\documentstyle[12pt]{article}

\newcommand{\Dlt}{\Delta}
\newcommand{\dlt}{\delta}
\newcommand{\br}{{\bf r}}
\newcommand{\bk}{{\bf k}}
\newcommand{\be}{\begin{equation}}
\newcommand{\ee}{\end{equation}}
\newcommand{\vp}{\varphi}
\newcommand{\om}{\omega}
\newcommand{\ep}{\varepsilon}

\begin{document}

\begin{center}
{\Large{\bf Number-of-particle fluctuations in systems with Bose-Einstein
condensate} \\ [5mm]
V.I. Yukalov} \\ [3mm]

{\it Institut f\"ur Theoretische Physik, \\
Freie Universit\"at Berlin, Arnimallee 14, D-14195 Berlin, Germany \\
and \\
Bogolubov Laboratory of Theoretical Physics, \\
Joint Institute for Nuclear Research, Dubna 141980, Russia}

\end{center}

\vskip 2cm

\begin{abstract}

Fluctuations of the number of particles for the dilute interacting gas with
Bose-Einstein condensate are considered. It is shown that in the Bogolubov 
theory these fluctuations are normal. The fluctuations of condensed as 
well as noncondensed particles are also normal both in canonical and grand
canonical ensembles.

\end{abstract}

\vskip 1cm


\vskip 2cm


\newpage

\section{Introduction}

Fluctuations in systems with Bose-Einstein condensate have long been a 
topic of special attention, as can be inferred from reviews [1--3]. This 
problem plays an important role in the general context of quantum statistical 
mechanics. Interest in this problem has quickened in the past years due to 
intensive experimental and theoretical investigations of Bose-Einstein 
condensed atomic gases (see reviews [4--8]).

The number-of-particle fluctuations in a uniform {\it ideal} Bose gas 
with Bose-Einstein condensate are known [1,2] to be anomalous, having the 
dispersion $\Dlt^2(\hat N)\sim N^2$,  with the power of $N$ larger than one. 
This anomalous behaviour comes from the fluctuations of the condensate, the 
corresponding dispersion being $\Dlt^2(\hat N_0)\sim N^2$. The fluctuations 
of the noncondensed particles are also anomalous, with the dispersion 
$\Dlt^2(\hat N_1)\sim N^{4/3}$. Such anomalous fluctuations manifest the
instability of the Bose-condensed ideal gas [3].

The problem of fluctuations in {\it interacting} Bose gases has been the arena
of numerous discussions, with controversial conclusions (see review [3]). In
the present paper, this problem is analysed in the frame of the Bogolubov
theory [9--11] and it is shown that no anomalous fluctuations arise, but all
fluctuations are normal.

\section{Number-of-particle fluctuations}

The measure of fluctuations for the number-of-particle operator $\hat N$ is
provided by the dispersion
\be
\label{1}
\Dlt^2(\hat N) \equiv \; <\hat N^2>\; - \; <\hat N>^2 \; ,
\ee
where the angle brackets mean, as usual, statistical averaging. Dispersion
(1) can be considered as an observable quantity, since it is directly connected
with the isothermal compressibility, sound velocity, and structural factor,
which are measurable quantities. In order to stress that the linkage between 
dispersion (1) and the observable quantities is {\it general} and {\it exact},
it is worth demonstrating how these relations can be rigorously derived.

For an equilibrium system of $N$ particles in volume $V$, with temperature
$T$, it is easy to check that
\be
\label{2}
\Dlt^2(\hat N) = k_B T \left ( 
\frac{\partial N}{\partial\mu}\right )_{TV} \; ,
\ee
where $k_B$ is the Boltzmann constant, $N\equiv<\hat N>$, and $\mu$ is chemical
potential. The Gibbs potential $G=G(T,P,N)$ is a function of temperature $T$,
pressure $P$, and the number of particles $N$. From the differential
$$
dG = - S\; dT + V\; dP + \mu\; dN \; ,
$$
taking into account that $G=\mu N$, one has
$$
N\; d\mu = - S\; dT + V\; dP \; ,
$$
where $S$ is entropy. The latter equality gives
$$
\left ( \frac{\partial\mu}{\partial N} \right )_{TV} = \frac{1}{\rho} \left (
\frac{\partial P}{\partial N} \right )_{TV} \; ,
$$
where $\rho\equiv N/V$ is density, or, equivalently,
$$
\left ( \frac{\partial N}{\partial\mu} \right )_{TV} =
\rho \left ( \frac{\partial N}{\partial P} \right )_{TV} \; .
$$
For the variables $N,\; P$, and $V$, because of the differential 
$$
dN =\left ( \frac{\partial N}{\partial P} \right )_V\; dP +
\left ( \frac{\partial N}{\partial V} \right )_P \; dV \; ,
$$
there exists the equality 
$$
\left ( \frac{\partial N}{\partial P} \right )_V 
\left ( \frac{\partial P}{\partial V} \right )_N
\left ( \frac{\partial V}{\partial N} \right )_P =  -1 \; .
$$
Using the Maxwell relation
$$
\left ( \frac{\partial V}{\partial N} \right )_{TP} =
\left ( \frac{\partial\mu}{\partial P} \right )_{TN}
$$
and the property
$$
\left ( \frac{\partial \mu}{\partial P} \right )_{TN}  =
\frac{1}{N} \left ( \frac{\partial G}{\partial P} \right )_{TN} =
\frac{1}{\rho} \; ,
$$
one gets
$$
\left ( \frac{\partial N}{\partial P} \right )_{TV} = -\rho
\left ( \frac{\partial V}{\partial P} \right )_{TN} \; .
$$
The definition of the isothermal compressibility
\be
\label{3}
\kappa_T \equiv -\; \frac{1}{V} 
\left ( \frac{\partial V}{\partial P} \right )_{TN} =
\frac{1}{\rho} \left ( \frac{\partial\rho}{\partial P} \right )_{TN} \; ,
\ee
owing to the above relations, can be rewritten as
\be
\label{4}
\kappa_T = \frac{1}{N} \left ( \frac{\partial N}{\partial P} \right )_{TV} =
\frac{1}{N\rho} \left ( \frac{\partial N}{\partial\mu} \right )_{TV} \; .
\ee
Comparing Eqs. (4) with (2) yields
\be
\label{5}
\kappa_T = \frac{\Dlt^2(\hat N)}{N\rho k_B T} \; .
\ee
In turn, the isothermal sound velocity $s$ can be expressed through the
compressibility as
\be
\label{6}
s^2 \equiv \frac{1}{m} \left ( \frac{\partial P}{\partial\rho} \right )_{TN} =
\frac{1}{m\rho\kappa_T} \; ,
\ee
where $m$ is particle mass.

On the other hand, representing the operator 
$\hat N\equiv\int\psi^\dagger(\br)\psi(\br)\; d\br$ through the field operators
$\psi^\dagger(\br)$ and $\psi(\br)$, from definition (1) it follows
\be
\label{7}
\Dlt^2(\hat N) = N + \int \rho(\br)\rho(\br') [g(\br,\br') -1 ]\; d\br
d\br' \; ,
\ee
which should be compared with the central structure factor
\be
\label{8}
S(0) =  1 + \frac{1}{N} \; \int \rho(\br) \rho(\br') [ g(\br,\br') - 1 ]\;
d\br \; d\br' \; ,
\ee
where $\rho(\br)\equiv<\psi^\dagger(\br)\psi(\br)>$, and the pair correlation
function is
\be
\label{9}
g(\br,\br') \equiv \;
\frac{<\psi^\dagger(\br)\psi^\dagger(\br')\psi(\br')\psi(\br)>}
{\rho(\br)\rho(\br')} \; .
\ee

In this way, there exist exact relations
\be
\label{10}
\Dlt^2(\hat N) = N k_B T \rho\kappa_T = N\; \frac{k_B T}{ms^2} =
NS(0) \; ,
\ee
valid for any equilibrium system, whether it is uniform or not. A stable
equilibrium system requires that its compressibility be positive and finite, 
$0<\kappa_T<\infty$. An infinite compressibility would mean that the system
immediately collapses or explodes. In a stable system, according to relations
(10), the sound velocity and the central structure factor are also finite and
positive. Thus, relations (10) tell us that in a stable equilibrium system
the dispersion $\Dlt^2(\hat N)$ must satisfy the {\it stability condition}
\be
\label{11}
0 < \frac{\Dlt^2(\hat N)}{N} < \infty \qquad (0<\kappa_T<\infty)
\ee
for any $N>0$, including the thermodynamic limit $N\rightarrow\infty$.
Fluctuations satisfying the stability condition (11) are termed {\it normal},
while if condition (11) is not valid, fluctuations are called {\it anomalous}.
Clearly, anomalous fluctuations imply instability. For instance, the ideal
uniform Bose gas is unstable, since it has anomalous fluctuations with
$\Dlt^2(\hat N)\sim N^2$ and a divergent compressibility $\kappa_T\sim N$.
Hydrodynamic equations for this gas with condensate are plagued by the
appearance of unbound solutions [12].

Now let us pass to systems experiencing Bose-Einstein condensation. At the
present time, a variety of trapped atomic gases is known to demonstrate this
phenomenon (see recent review [3]). Several types of molecules, produced by
means of Feshbach resonances, have been condensed. Bose-Einstein condensation
might also appear in a system of semiconductor biexcitons, consisting of two
electrons plus two holes [13,14].

In the presence of condensate, the operator of the total number of particles
is written as a sum
\be
\label{12}
\hat N = \hat N_0 + \hat N_1 \; ,
\ee
whose terms correspond to condensed $(\hat N_0)$ and noncondensed $(\hat N_1)$
particles. Then dispersion (1) takes the form
\be
\label{13}
\Dlt^2(\hat N) = \Dlt^2(\hat N_0) +
\Dlt^2(\hat N_1) + 2\; {\rm cov}(\hat N_0,\hat N_1) \; ,
\ee
in which the last term contains the covariance
\be
\label{14}
{\rm cov}(\hat N_0,\hat N_1) \equiv \frac{1}{2}\; <\hat N_0\hat N_1 +
\hat N_1 \hat N_0>\; - \; <\hat N_0><\hat N_1> \; .
\ee
Since $\hat N_0$ and $\hat N_1$ are usually defined as commuting operators,
covariance (14) reduces to
\be
\label{15}
{\rm cov}(\hat N_0,\hat N_1) =\; <\hat N_0\hat N_1>\; - \;
<\hat N_0><\hat N_1> \; .
\ee

By the stability condition (11), we know that $\Dlt^2(\hat N)\sim N$. Then
the fundamental question is: Could the fluctuations of either condensed or
noncondensed particles, or both, be anomalous, at the same time preserving 
the validity of the exact Eq. (13)? Recently there have appeared a number 
of papers stating that such fluctuations could be anomalous (see discussion
in review [3]). In the following section we consider an interacting homogeneous
system with Bose-Einstein condensate at low temperature and density, when the 
Bogolubov theory is applicable, and show that all fluctuations are normal.

\section{Dilute gas}

Current experiments with Bose-condensed atomic gases are typified by a rather 
low density, such that $\rho a_s^3\ll 1$, where $a_s$ is a scattering length. 
Atomic gases can be cooled down to very low temperatures, when practically all 
atoms are condensed, so that $|N_0 -N|\ll N$. Under these conditions, the 
Bogolubov theory [9--11] becomes applicable.

Let us start with the standard Hamiltonian for a uniform system of spinless
atoms,
\be
\label{16}
H = \int \psi^\dagger(\br) \left ( -\; \frac{\hbar^2}{2m}\;
\nabla^2 - \mu\right ) \psi(\br)\; d\br + 
\frac{1}{2}\; \int \psi^\dagger(\br)\psi^\dagger(\br')\Phi(\br-\br')
\psi(\br')\psi(\br)\; d\br d\br' \; ,
\ee
with an interaction potential $\Phi(\br)=\Phi(-\br)$. For dilute gas, 
the latter is usually modelled by the contact interaction 
$\Phi(\br)=(4\pi\hbar^2 a_s/m)\dlt(\br)$. However, for the sake of generality,
we keep the form $\Phi(\br)$. The chenical potential $\mu$ is included here
in order to compare the results, corresponding to the grand canonical ensemble,
with the original Bogolubov consideration [9--11] made in the frame of the
canonical ensemble.

Following the Bogolubov prescription let us separate the condensate by the 
shift in the field operator,
\be
\label{17}
\psi(\br) = \psi_0 +\psi_1(\br) \; ,
\ee
where the condensate operator $\psi_0$ does not depend on $\br$ owing to the
system uniformity. The field operators of noncondensed particles, $\psi_1(\br)$,
are assumed to possess the same Bose commutation relations as $\psi(\br)$. From
this it follows that the operator of condensed particles, $\psi_0$, commutes 
with all operators in the thermodynamic limit, when $N\rightarrow\infty$. The 
operators $\psi_0$ and $\psi_1$ are assumed to be orthogonal,
$$
\int \psi_0^\dagger \psi_1(\br) \; d\br = 0 \; .
$$
Hence, $\hat N=\hat N_0 +\hat N_1$ as in Eq. (12), with 
$$
\hat N_0 \equiv \psi_0^\dagger \psi_0 V \; , \qquad 
\hat N_1 \equiv \int \psi_1^\dagger(\br) \psi_1(\br) \; d\br \; .
$$

For a uniform system, the field operators can be expanded in Fourier series
$$
\psi_0 = \frac{a_0}{\sqrt{V}} \; , \qquad 
\psi_1(\br) = \sum_{k\neq 0} a_k \vp_k(\br)
$$
over the plane waves $\vp_k(\br)=e^{i\bk \cdot\br}/\sqrt{V}$. Assuming that 
the interaction potential can also be Fourier expanded, we have
$$
\Phi(\br) = \frac{1}{V} \; \sum_k \Phi_k e^{i\bk\cdot\br} \; , \qquad
\Phi_k = \int \Phi(\br) e^{-i\bk\cdot\br}\; d\br \; .
$$
For the number-of-particle operators, we get
$$
\hat N_0 = a_0^\dagger a_0 \; , \qquad 
\hat N_1 =\sum_{k\neq 0} a_k^\dagger a_k \; .
$$
Hamiltonian (16) is gauge invariant, hence the field operators are defined 
up to a global phase factor. Then one may chose a representation, where $a_0$ 
is self-adjoint, such that $a_0a_0=\hat N_0$ and $a_0^\dagger a_0^\dagger=\hat
N_0$. Actually, this choice is absolutely not principal and is accepted with 
the sole aim to simplify the following formulas.

After the Fourier transformation, Hamiltonian (16) acquires the form
\be
\label{18}
H = \sum_{n=0}^4 H^{(n)} \; ,
\ee
in which the terms are grouped according to the number of noncondensed-particle
operators. In the zeroth order,
$$
H^{(0)} = \frac{\hat N_0^2}{2V} \; \Phi_0 - \mu\hat N_0 \; ,
$$
where $\Phi_0\equiv \int\Phi(\br)d\br$. The first-order term vanishes, 
$H^{(1)}=0$. In the second order,
$$
H^{(2)} =\sum_{k\neq 0} \left ( \frac{\hbar^2 k^2}{2m} +
\frac{\hat N_0}{V}\; \Phi_0 - \mu\right ) a^\dagger_k a_k +
\frac{\hat N_0}{2V} \; \sum_{k\neq 0} \Phi_k \left ( 2a_k^\dagger a_k +
a_k^\dagger a_{-k}^\dagger + a_{-k} a_k \right ) \; ,
$$
where it is taken into account that $\Phi_{-k}=\Phi_k$ because of the symmetry
of $\Phi(\br)$. The third-order term is
$$
H^{(3)} = \frac{1}{V} \; \sum_{kq\neq 0} \Phi_q\left ( a_0^\dagger 
a_k^\dagger a_{k-q} a_q + a_q^\dagger a_{k-q}^\dagger a_k a_0 \right ) \; .
$$
Finally, the fourth-order term is written as
$$
H^{(4)} = \frac{1}{2V} \; \sum_{kpq\neq 0} \Phi_q a_k^\dagger a_p^\dagger
a_{p+q} a_{k-q} \; .
$$

Remembering that we consider an almost condensed system, where $N_0\approx N$,
the terms $H^{(3)}$ and $H^{(4)}$ can be treated as small perturbations to the
terms up to the second order. Limiting ourselves by the latter terms, we have
$$
H = \frac{\hat N_0^2}{2V}\; \Phi_0 - \mu \hat N_0 + \sum_{k\neq 0} 
\left (\frac{\hbar^2 k^2}{2m} + \frac{\hat N_0}{V}\; \Phi_0 -\mu \right ) 
a_k^\dagger a_k +
$$
\be
\label{19}
+ \frac{\hat N_0}{2V}\; \sum_{ k\neq 0} \Phi_k \left ( 2a_k^\dagger a_k +
a_k^\dagger a_{-k}^\dagger + a_{-k} a_k \right ) \; .
\ee
The contraction of the total fourth-order Hamiltonian (18) to its second-order
part (19) is the first Bogolubov approximation.

The next approximation is the replacement in Eq. (19) of the operator $\hat N_0$
by its average value $N_0$, which gives
\be
\label{20}
H =\frac{N_0^2}{2V}\; \Phi_0 - \mu N_0 + \sum_{k\neq 0} \om_k a_k^\dagger a_k
+ \frac{N_0}{2V}\; \sum_{k\neq 0} \Phi_k\left ( 2a_k^\dagger a_k +
a_k^\dagger a_{-k}^\dagger + a_{-k} a_k\right ) \; ,
\ee
where the notation
\be
\label{21}
\om_k \equiv \frac{\hbar^2 k^2}{2m} + \frac{N_0}{V}\; \Phi_0 - \mu
\ee
is used.

The Heisenberg equation of motion for $\psi_0$ is equivalent to the equation
$$
< \frac{\dlt H}{\dlt \hat N_0} > \; = 0 \; ,
$$
which yields
\be
\label{22}
\mu = \frac{N_0+N_1}{V}\; \Phi_0 + \frac{1}{2V} \; 
\sum_{k\neq 0}  \Phi_k\left ( 2a_k^\dagger a_k + a_k^\dagger a_{-k}^\dagger
+a_{-k} a_k\right ) \; .
\ee
To remain in the frame of the second-order approximation in Hamiltonian (20),
one has to use in Eq. (21) the zero-order part of the chemical potential (22),
setting there
\be
\label{23}
\mu = \frac{N_0}{V}\; \Phi_0 \; .
\ee

Another approximation is again based on the fact that $N_0\approx N$. Therefore
one can replace in Eq. (20) $N_0$ by $N$. This gives
\be
\label{24}
H = \frac{N^2}{2V}\; \Phi_0 + \sum_{k\neq 0} \om_k a_k^\dagger a_k -
\mu N + \frac{N}{2V} \; \sum_{k\neq 0} \Phi_k \left ( 2 a_k^\dagger a_k +
a_k^\dagger a_{-k}^\dagger + a_{-k} a_k \right ) \; .
\ee

Hamiltonian (24) is diagonalized by means of the Bogolubov canonical
transformation
$$
a_k = u_k b_k + v_{-k} b_{-k}^\dagger \; ,
$$
resulting in the diagonal form
\be
\label{25}
H = E_0 + \sum_{k\neq 0} \ep_k b_k^\dagger b_k - \mu N \; ,
\ee
in which the first term is the ground-state energy
$$
E_0 = \frac{1}{2}\; N\rho \Phi_0 + \frac{1}{2}\; \sum_{k\neq 0}
\left ( \ep_k - \om_k - \rho\Phi_k \right ) \; ,
$$
and the Bogolubov spectrum is
$$
\ep_k = \sqrt{2\rho \Phi_k\om_k +\om_k^2} \; .
$$
The latter, with the notation for the sound velocity
$$
c_k \equiv \sqrt{(\rho/m)\Phi_k} \; ,
$$
can be rewritten as 
$$
\ep_k =\sqrt{2m c_k^2 \om_k + \om_k^2} \; .
$$
Note that, due to Eqs. (21) and (23), the spectrum $\ep_k$ is gapless.
The coefficient functions of the Bogolubov transformation are given by the
equations
$$
u_k^2 =\frac{\sqrt{\ep_k^2+m^2c_k^4}+\ep_k}{2\ep_k} \; , \qquad
v_k^2 =\frac{\sqrt{\ep_k^2+m^2c_k^4}-\ep_k}{2\ep_k} \; .
$$
In the case of the contact interaction, one has $\Phi_k=4\pi\hbar^2 a_s/m$ and
$c_k\equiv c=(\hbar/m)\sqrt{4\pi\rho a_s}$.

The diagonal form (25) has been derived here starting with the Hamiltonian 
(16) corresponding to the grand canonical ensemble. The same Hamiltonian (25), 
up to the term $-\mu N$, was derived by Bogolubov [9--11] in the frame of the
canonical ensemble. Therefore all following calculations are actually identical
for the grand canonical as well as canonical ensembles.

A principal point must be stressed related to the transition from Eq. (19)
to Eq. (20), when the operator $\hat N_0$ is replaced by its average value
$N_0$. The equivalence of Hamiltonians (19) and (20) can be understood in
two senses. In the strong sense, the equality of operators (19) and (20)
requires that $\hat N_0=N_0$. This, however, is not compulsory. And the
equivalence of operators (19) and (20) can be understood in the weak sense,
implying the equality of all their matrix elements. The latter can be
reformulated as the equality on the average, so that the average values $<H>$ 
for both forms (19) and (20) be coinciding. As is evident from expressions (19)
and (20), their average values coincide then and only then, when the operator 
$\hat N_0$ is not correlated with the operators $a_k^\dagger a_k$ and
$a_{-k}a_k$. This means, in particular, the validity of the decoupling
\be
\label{26}
<\hat N_0\hat N_1>\; = \; <\hat N_0><\hat N_1>\; .
\ee

Taking account of Eq. (26), we notice that covariance (15) vanishes, 
${\rm cov}(\hat N_0,\hat N_1)=0$. As a result, dispersion (13) becomes
\be
\label{27}
\Dlt^2(\hat N) = \Dlt^2(\hat N_0) + \Dlt^2(\hat N_1) \; .
\ee
All terms here are non-negative. Hence, if at least one of the dispersions,
either $\Dlt^2(\hat N_0)$ or $\Dlt^2(\hat N_1)$, is anomalous, then the
left-hand side, $\Dlt^2(\hat N)$, is also anomalous. This, however, would
contradict the stability condition (11). Thus we come to an indispensable
conclusion: {\it In a stable equilibrium system, the fluctuations of condensed
as well as of noncondensed atoms must be normal}.

If all fluctuations have to be normal, then how could one explain the 
appearance of anomalous fluctuations in a number of recent calculations (see
review [3]) accomplished on the basis of the same Bogolubov approximations?
To answer this question, let us recollect how such anomalous fluctuations
arise. The standard origin of their appearance in calculations is as follows.
One considers the dispersion $\Dlt^2(\hat N_1)$, containing the four-operator
terms $<a_k^\dagger a_ka_q^\dagger a_q>$ or $<b_k^\dagger b_kb_q^\dagger b_q>$.
These are decoupled as $<b_k^\dagger b_k><b_q^\dagger b_q>+<b_k^\dagger b_q>
<b_kb_q^\dagger>$. Calculating $\Dlt^2(\hat N_1)$, one meets the integral
$\Dlt^2(\hat N_1)\sim N\int dk/k^2$. To avoid here the infrared divergence,
one can replace the integral by a sum over the discretized spectrum of
collective excitations [1,15]. This yields $\Dlt^2(\hat N_1)\sim N^{4/3}$.
Another way could be by limiting the integration from below by $k_{min}\approx
1/L$, with $L\sim N^{1/3}$. Then again $\Dlt^2(\hat N_1)\sim N^{4/3}$ in either
canonical or grand canonical ensemble. This means, according to equality (27), 
that the dispersion $\Dlt^2(\hat N)\sim N^{4/3}$ is anomalous, hence the
system is unstable.

In order to stress that the same type of the anomalous dispersion 
$\Dlt^2(\hat N_1)$ arises in the grand canonical as well as in the canonical
ensembles, let us consider $\Dlt^2(\hat N_1)$ in the grand canonical ensemble,
when it can be represented as
$$
\Dlt^2(\hat N_1) = k_B T\; \frac{\partial}{\partial\mu}\;
<\hat N_1>\; - \; {\rm cov}(\hat N_0,\hat N_1) \; .
$$
Since in the Bogolubov approximation ${\rm cov}(\hat N_0,\hat N_1)=0$, one 
has
$$
\Dlt^2(\hat N_1) = k_B T\; \frac{\partial}{\partial\mu}\;
<\hat N_1>\; .
$$
Keeping in mind that $v_{-k}=v_k$, the operator $\hat N_1$ can be written 
as
$$
\hat N_1 =\sum_{k\neq 0} \left [ \left ( u_k^2 + v_k^2 \right )
b_k^\dagger b_k + v_k^2 + u_k v_k \left ( b_k^\dagger b_{-k}^\dagger +
b_{-k}^\dagger b_k \right ) \right ] \; .
$$
From here one gets
$$
\frac{<\hat N_1>}{V} = \frac{1}{2}\; \int
\left [ \frac{\sqrt{\ep_k^2+m^2c_k^4}}{\ep_k}\; {\rm coth}\left (
\frac{\ep_k}{2k_BT}\right ) - 1 \right ]\;
\frac{d\bk}{(2\pi)^3} \; .
$$
For the case of the contact interaction, when $c_k=c$, and involving the
relation
$$
\om_k =\sqrt{\ep_k^2 + m^2 c^4} - mc^2 \; ,
$$
we come to the form
$$
\frac{<\hat N_1>}{V} = \frac{\sqrt{2}}{(2\pi)^2} \left (
\frac{mc}{\hbar}\right )^3 \; \int_{x_0}^\infty\; \left [
\sqrt{1+x^2} -1 -\dlt \right ]^{1/2} \left [ {\rm coth} \left (
\frac{mc^2}{2k_BT}\; x\right ) -
\frac{x}{\sqrt{1+x^2}} \right ] \; dx \; ,
$$
in which
$$
x_0\equiv \sqrt{\dlt(2+\dlt)} \; , \qquad 
\dlt\equiv \frac{\rho_0\Phi_0-\mu}{mc^2} \; .
$$
At zero temperature, one has the known result
$$
\frac{<\hat N_1>}{N} = \frac{8}{3\sqrt{\pi}}\; \sqrt{\rho a_s^3} \qquad
(T=0) \; .
$$
For any nonzero temperatures, taking the derivative 
$\partial<\hat N_1>/\partial\mu$ and setting $\mu=\rho_0\Phi_0$, one recovers
the same anomalous behaviour $\Dlt^2(\hat N_1)\sim N \int dx/x^2\sim N^{4/3}$.

However this anomaly is spurious and arises owing to a not self-consistent
calculational procedure. Really, the Bogolubov approach is based on the
Hamiltonian (19) of the second-order with respect to noncondensed-particle
operators $a_k$. All higher-order terms have been omitted from the initial
Hamiltonian (18). Hence, such higher-order terms should also be omitted
from the calculations of any other physical quantities. Calculating the 
fourth-order expression $<\hat N_1^2>$ with the second-order approximation
is not self-consistent, hence, is not correct.

The total dispersion $\Dlt^2(\hat N)$ can be found from Eq. (7). In order 
to be self-consistent in defining the pair correlation function $g(\br-\br')=
g(\br,\br')$, one has to omit in the latter all terms of the order higher 
than two. Then one finds
\be
\label{28}
g(\br) = 1 + \frac{2}{\rho} \int \left ( <a_k^\dagger a_k> \; + \;
<a_k a_{-k}> \right ) e^{i\bk\cdot\br} \;
\frac{d\bk}{(2\pi)^3} \; .
\ee
From here,
$$
\int [ g(\br)-1]\; d\br = \frac{2}{\rho}\; \lim_{k\rightarrow 0} \left (
<a_k^\dagger a_k>\; + \; <a_k a_{-k}>\right ) \; .
$$
This gives for the structural factor (8)
\be
\label{29}
S(0) = \frac{k_B T}{mc^2} \qquad 
\left ( c\equiv \sqrt{\frac{\rho}{m}\; \Phi_0} \right ) \; .
\ee
And we find the dispersion (10) as
\be
\label{30}
\Dlt^2(\hat N) = N\; \frac{k_BT}{mc^2} \; ,
\ee
which is, of course, normal. If in the pair correlation function (28) we 
would retain the terms of the orders higher than the second with respect to 
$a_k$, we would again get an anomalous dispersion. This, however, would not 
involve any senseful physics, but would simply mean the inconsistency of 
the calculations.

The dispersion $\Dlt^2(\hat N_1)$, being a fourth-order expression 
with respect to $a_k$, is not a well-defined quantity in the frame of the 
second-order Bogolubov theory. For its definition, it requires to invoke 
additional assumptions. Thus, we may follow the Bogolubov approximation,
treating $\hat N_0=N_0$ as a classical quantity, because of which 
$\Dlt^2(\hat N_0)=0$. Then in the grand canonical ensemble,
$$
\Dlt^2(\hat N_1) = \Dlt^2(\hat N) = N\; \frac{k_BT}{mc^2} \qquad
(grand) \; .
$$
In the canonical ensemble, when $\hat N=N$ is fixed, we get
$$
\Dlt^2(\hat N_0) = \Dlt^2(\hat N_1) = N\; \frac{k_BT}{mc^2} \qquad
(canonical) \; ,
$$
which results from the fact that $\Dlt^2(\hat N_1)$ is the same in both
ensembles.

Passing to the ideal gas, one should consider the limit 
$\Phi(\br)\rightarrow 0$, that is, $c\rightarrow 0$. Then $\Dlt^2(\hat N)$
diverges, as it should be for the ideal gas, which is unstable. The character
of this divergence can be estimated as follows. The vanishing of $\Phi(\br)$
can be accepted as being analogous to its diminishing at large distance, where
it has the standard Lennard-Jones behaviour $\Phi(\br)\sim 1/r^6$. In other
words, $\Phi(\br)\sim 1/N^2$. Consequently, $\Phi_0\sim 1/N$, hence
$c\sim 1/\sqrt{N}$. Then dispersion (30) diverges as $\Dlt^2(\hat N)\sim N^2$,
which is typical of the ideal uniform Bose gas [1,2]. Contrary to the unstable
Bose-condensed ideal gas, the interacting gas is stable, possessing always only
normal fluctuations.

The conclusion on the absence of anomalous fluctuations can be generalized for
any stable equilibrium systems by rigorously proving the following theorem.
If an observable quantity is represented as a sum $\hat A+\hat B$ of two linearly
independent terms, then the total dispersion
$$
\Dlt^2(\hat A + \hat B) = \Dlt^2(\hat A) +\Dlt^2(\hat B) + 
2\;{\rm cov}(\hat A,\hat B) \; ,
$$
with the covariance
$$
{\rm cov}(\hat A,\hat B) \equiv \frac{1}{2}\; <\hat A\hat B +
\hat B \hat A>\; - \; <\hat A><\hat B> \; ,
$$
is normal then and only then, when both partial dispersions $\Dlt^2(\hat A)$
and $\Dlt^2(\hat B)$ are normal. The proof will be presented in a separate
publication.

\vskip 5mm

{\bf Acknowledgement}. I am grateful for useful discussions to L.P. Pitaevskii
and for helpful comments to Z. Idziaszek and E.P. Yukalova. The Mercator 
Professorship of the German Research Foundation is appreciated.

\end{document}